\documentclass[aps, reprint, superscriptaddress, prl, footinbib]{revtex4-1}
\usepackage{amsmath}
\usepackage{amsfonts}
\usepackage{amssymb}
\usepackage{graphicx}
\usepackage{color}
\usepackage{natbib}

\begin{document}
	
\title{Phase-controlled bistability of a dark soliton train in a polariton fluid}

\author{V. Goblot}
\email{valentin.goblot@lpn.cnrs.fr}
\affiliation{Centre de Nanosciences et de Nanotechnologies, CNRS, Univ. Paris-Sud, Universit\'e Paris-Saclay, C2N - Marcoussis, 91460 Marcoussis, France}

\author{H. S. Nguyen}
\affiliation{Institut de Nanotechnologies de Lyon, Ecole Centrale de Lyon, CNRS (UMR 5270), 69134 Ecully, France}

\author{I. Carusotto}
\affiliation{INO-CNR BEC Center and Dipartimento di Fisica, Universit\`a di Trento, I-38123 Povo, Italy}

\author{E. Galopin}
\author{A. Lema\^itre}
\author{I. Sagnes}
\author{A. Amo}
\affiliation{Centre de Nanosciences et de Nanotechnologies, CNRS, Univ. Paris-Sud, Universit\'e Paris-Saclay, C2N - Marcoussis, 91460 Marcoussis, France}

\author{J. Bloch}
\affiliation{Centre de Nanosciences et de Nanotechnologies, CNRS, Univ. Paris-Sud, Universit\'e Paris-Saclay, C2N - Marcoussis, 91460 Marcoussis, France}
\affiliation{D\'epartement de Physique, Ecole Polytechnique, Universit\'e Paris Saclay, F-91128 Palaiseau Cedex, France}

\date{\today}

\begin{abstract}
We use a one-dimensional polariton fluid in a semiconductor microcavity to explore the rich nonlinear dynamics of counter-propagating interacting Bose fluids. The intrinsically driven-dissipative nature of the polariton fluid allows to use resonant pumping to impose a phase twist across the fluid. When the polariton-polariton interaction energy becomes comparable to the kinetic energy, linear interference fringes transform into a train of solitons. A novel type of bistable behavior controlled by the phase twist across the fluid is experimentally evidenced.
\end{abstract}



\maketitle

Dark solitons are among the fundamental nonlinear collective excitations of one-dimensional (1D) quantum degenerate fluids with positive mass and repulsive interactions. They are characterized by a dip in a uniform background density and a jump in the macroscopic phase across it. The shape and size of the dip is given by the interplay of mass and nonlinearity. Because of the universality of the mechanisms necessary to their formation, dark solitons have been observed in a wide variety of systems ranging from Bose-Einstein condensates of cold atoms~\cite{Burger1999, Denschlag2000, Engels2007b}, optical fibers~\cite{Weiner1988}, to thin magnetic films~\cite{Chen1993}. Interestingly, dark solitons have also been observed in nonlinear open-dissipative systems, in particular, in semiconductor microcavities~\cite{Larionova2008, Amo2011, Grosso2011, Hivet2012a} and are attracting great interest in view of photonic applications~\cite{Ackemann2009}.

Semiconductor microcavities have recently appeared as an excellent platform to study the nonlinear dynamics of interacting Bose fluids in a photonic context~\cite{Carusotto2013}. Their elementary excitations are exciton-polaritons, bosonic quasiparticles arising from the strong coupling between quantum well excitons and photons confined in the microcavity. While their excitonic component provides significant repulsive interactions, the fast escape of photons out of the microcavity makes polaritons an intrinsically open-dissipative system, requiring continuous wave pumping to achieve a steady state.
A number of quantum fluid effects have been studied in semiconductor microcavities, including superfluidity~\cite{Amo2008}, diffusive Goldstone modes~\cite{Ballarini2009a}, Bogoliubov excitation spectrum~\cite{Kohnle2011a}, solitary bright waves~\cite{Amo2009, Sich2012}, and the hydrodynamic nucleation of quantized vortices~\cite{Nardin2011, Sanvitto2011} and dark solitons \cite{Amo2011, Grosso2011}.

In addition to the possibility of in-situ and time-resolved imaging of the fluid dynamics, a remarkable feature of driven-dissipative systems is that a resonant drive allows setting the local phase of the wavefunction~\cite{Carusotto2013}. It is then possible to externally manipulate the boundary conditions and impose a controlled phase pattern across a polariton fluid. This was first explored in a two-dimensional polariton condensate in which a spatial vortex phase profile was imposed on the polariton field, resulting in persistent currents with high orbital momentum~\cite{Sanvitto2009}. This technique opens up a new world for the exploration of the elementary excitations of polariton quantum fluids. In particular, it has been proposed that by imposing a phase twist across the fluid via the external pumping, the superfluid fraction could be measured~\cite{Janot2013}, different Josephson dynamical regimes could be addressed~\cite{Sarchi2008,Shelykh2008}, and the controlled nucleation of dark solitons could be implemented~\cite{Petrov2013}.

In this Letter, we report on the study of counter-propagating interacting polariton fluids resonantly excited in a 1D semiconductor cavity. At high excitation power, polariton-polariton interactions are responsible for the self-organization of a dark soliton train, which is directly evidenced by spatial imaging of the 1D channel. When scanning the excitation power, the abrupt disappearance of solitons reflects the discrete nature of these nonlinear excitations. Interestingly, varying the phase difference between the two pumping beams, we are able to impose a phase twist across the fluid which controls not only the position of the soliton train, but also the parity of their number. A novel type of bistable behavior appears when scanning the phase twist up and down, at constant power.

\begin{figure}[t]
	\includegraphics[width=\linewidth]{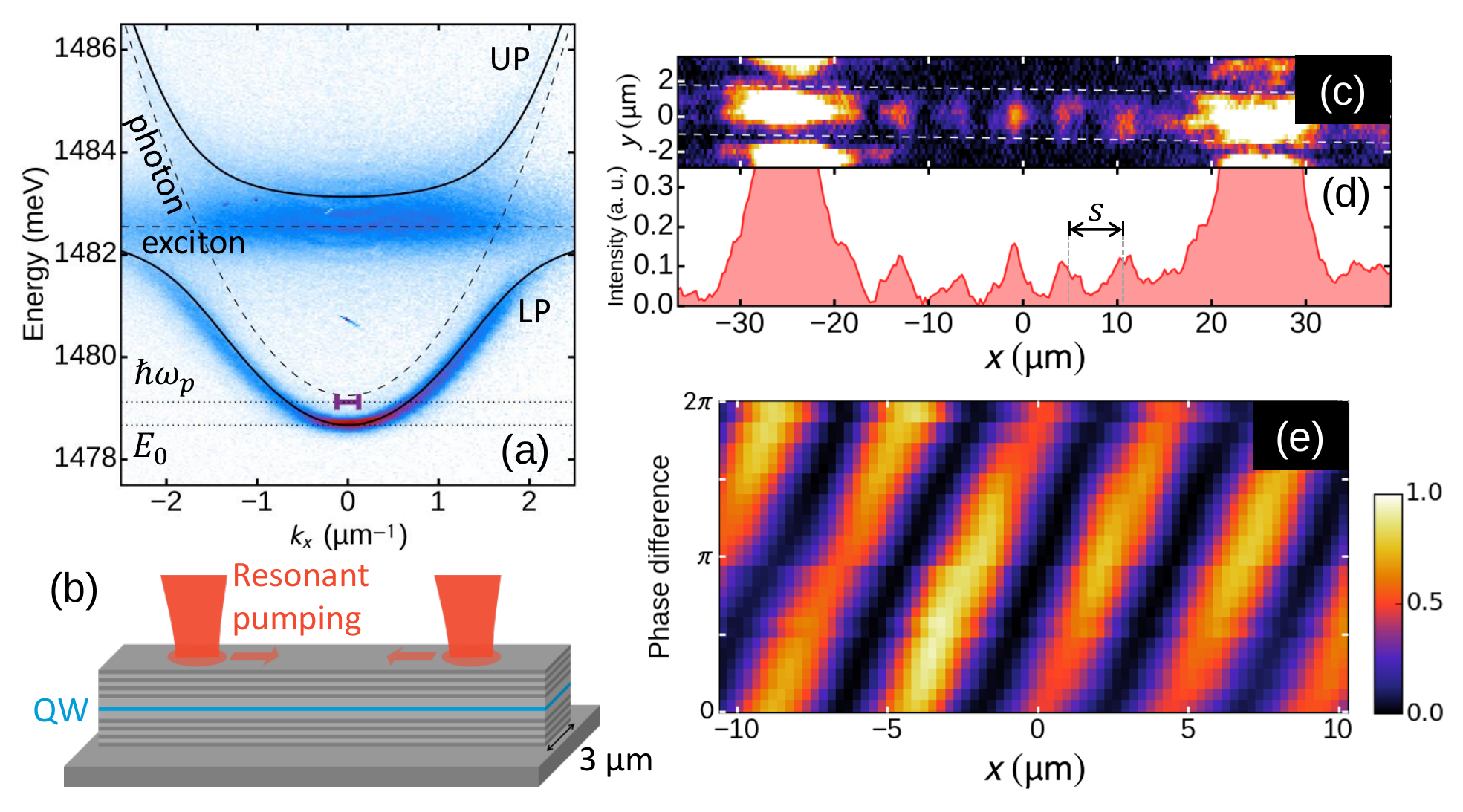}
	\caption{\label{fig1} (a) Far-field photoluminescence measured under non-resonant pumping. (Solid lines) theoretical fits of the lower and upper polariton branches and (dashed lines) bare exciton and photon energy. The horizontal segment shows the energy and width for the resonant excitation conditions. (b) Sketch of the experimental configuration. (c) Spatially resolved emission measured along the wire in the linear regime, for  $P = 8\ \mathrm{mW}$, $\Delta E = 0.27 \ \mathrm{meV}$, $d = 50 \ \mathrm{\mu m}$ and $\Delta \varphi \approx 0$. Dotted lines indicate the wire edges. The color scale is saturated in regions under the spots. (d) Measured intensity profile integrated in the transverse direction. (e) Intensity profile measured along the wire as a function of $\Delta \varphi$ for similar pumping parameters, well in the linear regime.
	}
\end{figure}

Our sample, grown by molecular beam epitaxy, consists of a $\lambda$ GaAs cavity surrounded by two $\mathrm{Ga_{0.9}Al_{0.1}As / Ga_{0.05}Al_{0.95}As}$ Bragg mirrors with 26 (30) pairs in the top (bottom) mirror. A single 8 nm $\mathrm{In_{0.95}Ga_{0.05}As}$ quantum well is inserted at the cavity center. The Rabi splitting resulting from the exciton-photon strong coupling amounts to $3.5 \ \mathrm{meV}$. Electron beam lithography and dry etching are used to fabricate photonic wires of $3 \ \mathrm{\mu m}$ width and $200 \ \mathrm{\mu m}$ length (Fig.~\ref{fig1}(b)). The experiments are performed at a temperature of 10 K. The photoluminescence is collected in transmission geometry through the sample back, and real- and momentum-space emission is imaged on a CCD camera coupled to a spectrometer. A polarizer selects the emission polarized along the wire.

First, the polariton dispersion in the wire has been characterized by non-resonant photoluminescence, using a cw single-mode Ti:sapphire laser. Figure~\ref{fig1}(a) shows the momentum-space emission, evidencing the lower and upper polariton 1D sub-bands. We deduce an exciton-photon detuning $\delta = E_C(k=0) - E_X(k=0) \ \approx -3.5 \ \mathrm{meV}$, where $E_C(k)$ is the bare photon energy and $E_X(k)$ the bare exciton energy. Close to $k=0$, the lower polariton branch can be approximated by a parabola, $E(k) = E_0 + \hbar^2 k^2 / 2 m$, where $m = 4 \times 10 ^{-5} \ m_{e}$ is the polariton effective mass and $m_e$ the free electron mass.

The focus of this Letter is to investigate the dynamics of a pair of counter-flowing polariton fluids. To create them, we use a resonant cw laser split in two separate beams, linearly polarized along the wire and focused at normal incidence onto two $8 \ \mathrm{\mu m}$ diameter spots separated by a distance $d$. The laser energy $\hbar \omega_{p}$ is blueshifted by $\Delta E = \hbar \omega_{p} - E_0$ with respect to the lower polariton energy $E_0$ at $k=0$ (see Fig.~\ref{fig1}(a)). The phase difference $\Delta \varphi$ between the two beams can be varied using a delay stage controlled by a piezoelectric actuator added to the path of one of the excitation beams.

Figure~\ref{fig1}(c) shows the polariton emission, spatially resolved along the wire, measured for a low pumping power $P = 8\ \mathrm{mW}$ well in the linear regime. The saturated bright regions correspond to the excitation spots positions, and the bright regions outside of the wire, above and below the spots, correspond to laser light scattered by the wire edges, and are thus not relevant. Even though the excitation spots are at normal incidence, their finite angular aperture allows injecting polaritons with wavevectors $k_f = \pm\sqrt{2 m \Delta E} / \hbar = \pm 0.53 \mathrm{\mu m^{-1}}$ \cite{Carusotto2013}. Between the two excitation spots, we observe a regular fringe pattern with a spacing of $s = 6.0 \ \mathrm{\mu m}=\pi / k_f$, arising from interference of the two counter-propagating polariton waves (a sinusoidal fit of the fringe pattern is shown in the Supplemental Material~\cite{Supplementary}). The position of the fringe pattern is determined by the boundary conditions imposed by the excitation spots, namely the distance between them and their phase difference $\Delta \varphi$. When $\Delta \varphi$ is scanned, we observe a continuous spatial displacement of the interference pattern (see Fig.~\ref{fig1}(e)), a behavior characteristic of the linear regime.

\begin{figure}[t]
	\includegraphics[width=\linewidth]{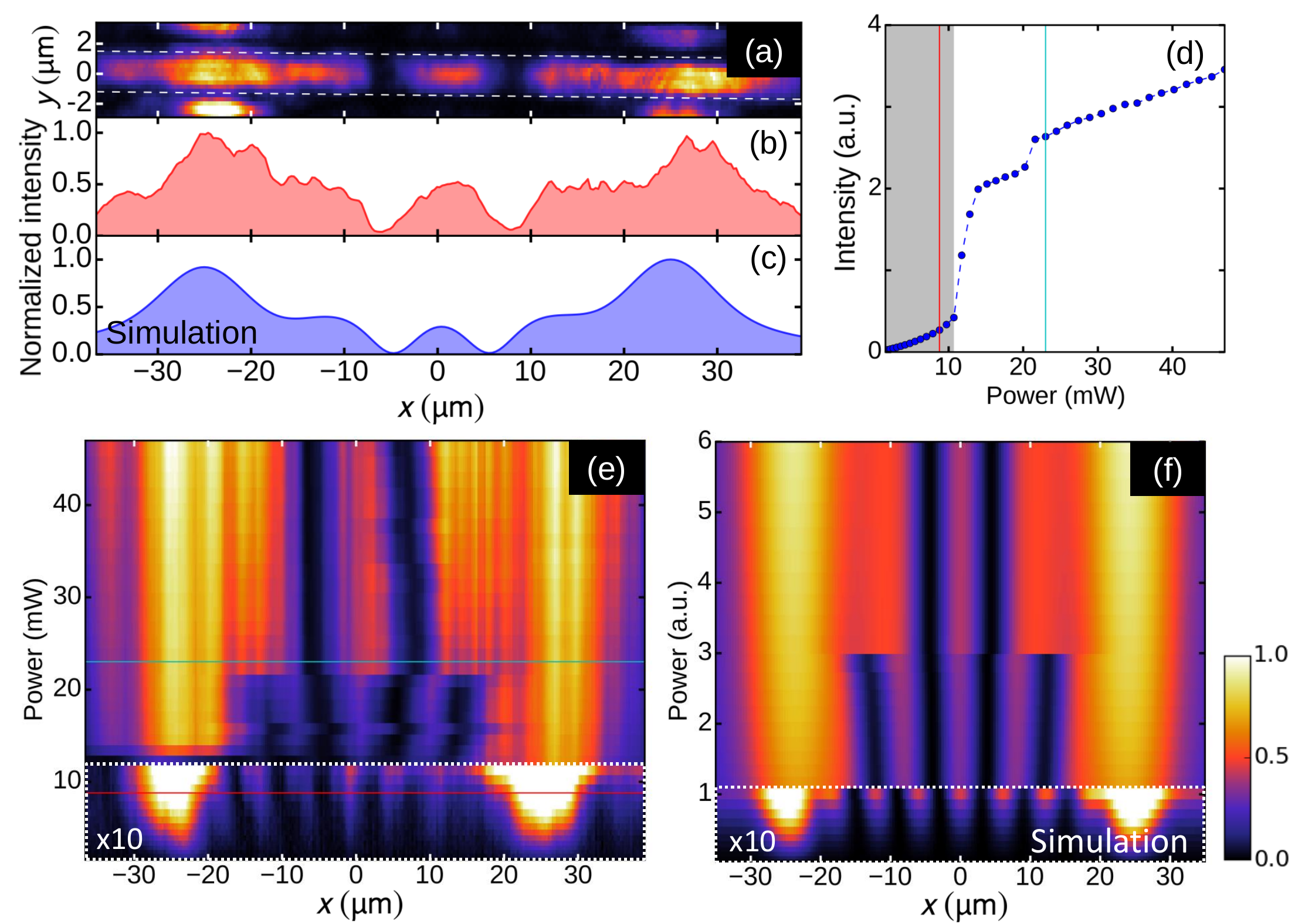}
	\caption{\label{fig2} (a) Spatially resolved emission measured along the wire for $P = 23\ \mathrm{mW}$ ($\Delta E = 0.27 \ \mathrm{meV}$, $d = 50 \ \mathrm{\mu m}$, $\Delta \varphi \approx 0$), and (b) intensity profile integrated over the transverse direction. (c) Corresponding calculated emission profile. (d) Total measured emission intensity (integrated along both the transverse and longitudinal directions) as a function of pump power. The shaded gray region corresponds to the linear regime. (e) Measured --(f) calculated-- emission profile when scanning the power up (the low power data have been amplified by a factor 10 for clarity). The horizontal red (resp. blue) line corresponds to the profile shown in Fig.~\ref{fig1}(d) (resp.~\ref{fig2}(b)).
	}
\end{figure}

Superfluidity and the nucleation of dark solitons are features of quantum fluids showing up when the interparticle interaction energy is comparable to the kinetic energy. In our system, the interaction energy is $\hbar g n$, where $g$ is the polariton-polariton interaction constant and $n$ the polariton density, controlled by the excitation power. When the latter is ramped up into the nonlinear regime, a first threshold is observed when the blueshift due to polariton-polariton interactions under the pump spots equals $\Delta E$, resulting in an abrupt increase of the polariton density (see Fig. 2(d)). When the power is further ramped up, a second threshold is observed with a change of the spatial pattern. A similar threshold behavior has been reported in a 1D polariton fluid in a configuration in which polaritons are excited by a single beam and reflected by an external potential ~\cite{Nguyen2015}.

A typical emission pattern above the second threshold is shown in Fig.~\ref{fig2}(a) for $\Delta \varphi \approx 0$. It strongly differs from the linear case [Fig.~\ref{fig1}(c)]: two density dips, dropping almost to zero, are visible in an otherwise almost constant high density profile.
Those dips are identified as dark solitons --nonlinear collective excitations of the fluid. They are well fitted by the characteristic hyperbolic tangent function, and the characteristic $\pi$ phase jump of the wavefunction across the soliton was experimentally confirmed by measuring the interference between the polariton emission and a constant-phase reference beam~\cite{Supplementary}.

Figure~\ref{fig2}(e) reveals that the number of solitons depends on the excitation power. Directly above the first threshold at $P_{th} = 12 \ \mathrm{mW}$, four solitons are present in the region between the spots. Further increasing the excitation power, we observe at $P = 21 \ \mathrm{mW}$ the abrupt expulsion of two solitons so that only two of them remain. Interestingly, the polariton density between the spots and outside of the dark solitons is almost independent of the pumping power. Notice that the observed expulsion of two solitons, replaced by regions of high polariton density, is responsible for the small jump in total emitted intensity that is visible at the second threshold (see Fig.~\ref{fig2}(d)). Throughout the whole power scan in Fig.~\ref{fig2}(d), the number of solitons remains even because of the symmetry of the excitation conditions. Indeed, since we impose $\Delta \varphi \approx 0$, the polariton wave function must remain symmetric, implying an even number of solitons.

To reproduce these experimental observations, we solve a 1D Gross-Pitaevskii equation that includes pump and loss terms, and consider only the lower polariton branch~\cite{Carusotto2004}. The evolution of the polariton wave function $\Psi (x)$ is given by:
\begin{align}
i \hbar \frac{\partial \Psi (x,t)}{\partial t} =& \left[ E_0 - \frac{\hbar^2}{2 m} \frac{\partial^2 \Psi (x,t)}{{\partial x}^2 } + \hbar g |\Psi (x)|^2 \right] \Psi (x,t) \nonumber \\
 &-i \frac{\hbar \gamma}{2} \Psi (x,t) + i F(x) e^{-i \omega_p t}
\label{eq:1}
\end{align}
where $\gamma$ is the polariton decay rate. $F(x)\,=\,F_0\, f(x)$, with $|F_0|^2$ being proportional to the total power of the coherent drive, and $f(x)$ is a complex function describing the spatial profile and the relative phase of the pump beams. The steady-state solutions of the equation are obtained numerically for the experimentally measured linewidth $\hbar \gamma = 47 \ \mathrm{\mu eV}$, $E_0 = 1478.57 \ \mathrm{meV}$, $\Delta E = 0.27 \ \mathrm{meV}$ and gaussian spots of width $w = 8 \ \mathrm{\mu m}$ separated by $d = 50 \ \mathrm{\mu m}$. We take an interaction constant $\hbar g = 0.3 \ \mathrm{\mu eV . \mu m}$~\cite{Ferrier2011}. Figure~\ref{fig2}(f) shows the calculated polariton density $|\Psi (x)|^2$ as a function of the excitation power $|F_0|^2$. The calculations perfectly reproduce the low power interferences and the abrupt transition to the nonlinear regime resulting in the nucleation of four dark solitons and, at higher power, two dark solitons [Fig.~\ref{fig2}(c)].

The nucleation of solitons in the nonlinear regime, and the abrupt change in their number when increasing the excitation power can be intuitively understood from the hydrodynamics of the polariton flow. In the steady state, in the central region far from the excitation spots, the real part of Eq.~(\ref{eq:1}) multiplied by $\Psi^*(x)$ can be written as a "local" energy conservation law as follows:
\begin{equation}
\hbar \omega_p = E_0 - \frac{\hbar^2}{2 m} \frac{\mathrm{Re} (\Psi^* \nabla ^2 \Psi)(x)}{n(x)} + \hbar g n(x)
\label{eq:3}
\end{equation}

The imaginary part of the steady state equation gives a continuity equation that accounts for the losses due to the finite polariton lifetime and the injection from the pumping beams. Equation~(\ref{eq:3}) shows that the energy per polariton is fixed by $\omega_p$. Thus, locally, $\hbar \omega_p$ must be equal to the sum of three terms: the single-polariton energy $E_0$ at $k=0$; a kinetic term ($- \frac{\hbar^2}{2 m} \frac{\mathrm{Re} (\Psi^* \nabla ^2 \Psi)}{n(x)}$); and a polariton-polariton interaction term ($ \hbar g n(x)$).

The specific dark soliton profile at a given pump power is a result of the local interplay between the kinetic and interaction terms. In the core of a soliton, where the density is low and its second order derivative is high, the kinetic term dominates over interactions, while it is the opposite in the high density regions far from the core. At pump densities just above the first nonlinear threshold, the polariton flow from the pump spots towards the central region contains a high kinetic energy that needs to be accommodated in the form of a large number of solitons, four in the case depicted in Fig.~\ref{fig2}(e) in the $12-21\ \mathrm{mW}$ range. When the excitation power is further increased, the higher density in the wire results in an increase of interactions. In the balance established by Eq.~(\ref{eq:3}), a higher weight of the interaction term must be accompanied by a decrease of the kinetic term, resulting in the expulsion of solitons.
The results of the numerical simulations [Fig.~\ref{fig2}(c),(f)] reproduce quantitatively the features observed in the experiment: at low pump intensities, there is just a linear interference whereas when interactions become significant, the sinusoid transforms into a soliton train, more precisely an elliptic function shape~\cite{Supplementary}, as first discussed in~\cite{Marburger1978} and~\cite{Carr2000}.

\begin{figure}[t]
	\includegraphics[width=\linewidth]{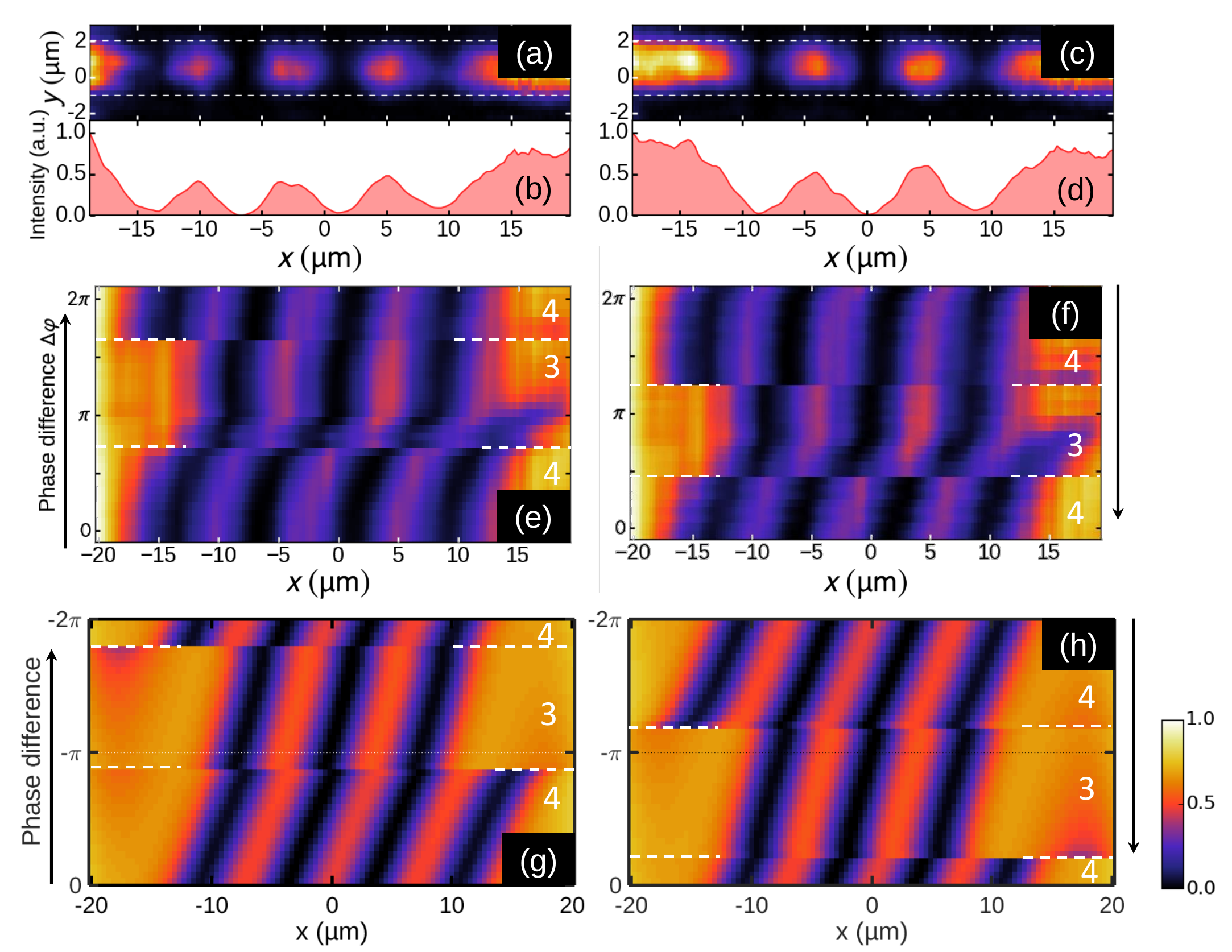}
	\caption{\label{fig3} (a), (c) Spatially resolved emission for $P = 57 \ \mathrm{mW}$, $\Delta E = 0.37 \ \mathrm{meV}$, $d = 50 \ \mathrm{\mu m}$ and: (a) $\Delta \varphi \approx 0$; (c) $\Delta \varphi \approx \pi$. (b), (d) Corresponding intensity profiles integrated over the transverse direction. (e), (f) Measured --(g), (h) calculated-- intensity profiles for increasing (e),(g) and decreasing (f),(h) phase difference $\Delta \varphi$ between the spots. White dotted lines indicate the value of $\Delta \varphi$ for which a soliton is expelled or generated. The measured number of solitons is indicated in white.
	}
\end{figure}

We now address bistability in the wire, a well-established behavior displayed by nonlinear dissipative systems as a function of driving intensity~\cite{Boyd2008, Baas2004b, Carusotto2013}. Usually this effect is observed in a configuration where the polariton field is frozen in a single mode. When, as in the present situation, multi-mode polariton fluids are considered, the complex spatial dynamics is expected to give rise to conceptually different bistability or even multistability effects~\cite{Paraiso2011, Rodriguez2016, Ouellet-Plamondon2016}. As a first example, we notice that the abrupt change in soliton number occurs at different excitation powers when the power is scanned downward than when the power is ramped up~\cite{Supplementary}: two different soliton patterns can thus be observed for the same excitation power.

An even more intriguing bistable behavior occurs when the excitation power is kept constant while scanning the phase twist $\Delta \varphi$ across the polariton fluid imposed by the excitation lasers.
Figure~\ref{fig3}(a)-(f) shows the polariton density profiles for a fixed excitation power and different values of $\Delta \varphi$.
For $\Delta \varphi = 0$ [Fig.~\ref{fig3}(a)] a symmetric profile is observed with four solitons. On the contrary, when $\Delta \varphi = \pi$ [Fig.~\ref{fig3}(c)], an antisymmetric profile is measured, with only three solitons, consistent with the antisymmetric boundary conditions.
The transition between the two situations takes place abruptly when scanning $\Delta \varphi$, as shown in Fig.~\ref{fig3}(e) (white dashed lines), attesting the nonlinear character of the fluid (in contrast to the smooth rigid motion of fringes in the linear regime that is visible in Fig.~\ref{fig1}(e)).
This transition can be understood in a similar way to the case of Fig.~\ref{fig2}, where a scan in power induces a change in interaction energy.
In the present situation, the phase twist results in a change in kinetic energy across the fluid, which is accommodated via the expulsion or addition of a soliton to the fluid pattern. When approaching $\Delta \varphi =\pi$, the choice between the expulsion and the inclusion of a soliton is settled by the most stable solution at the considered excitation power. 

\begin{figure}[t]
	\includegraphics[width=0.9\linewidth]{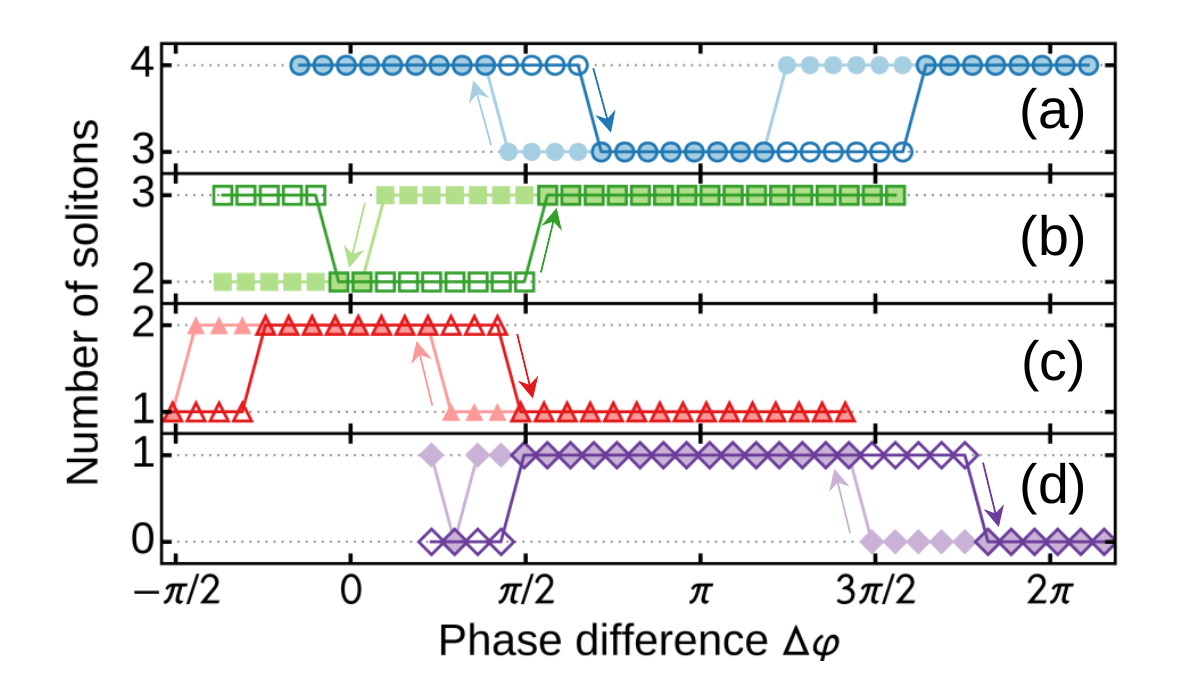}
	\caption{\label{fig4} (a)-(d) Number of solitons measured when scanning $\Delta\varphi$ up (empty symbols) and down (full symbols). (a) Same parameters as in Fig.~\ref{fig3}(e),(f); (b) $\Delta E = 0.21 \ \mathrm{meV}$, $P = 42 \ \mathrm{mW}$, $d = 60 \ \mathrm{\mu m}$; (c) $\Delta E = 0.35 \ \mathrm{meV}$, $P = 90 \ \mathrm{mW}$, $d = 40 \ \mathrm{\mu m}$; (d) $\Delta E = 0.20 \ \mathrm{meV} $, $P = 103 \ \mathrm{mW} $, $d = 40 \ \mathrm{\mu m}$. The fluctuations due to phase noise in the experimental setup are estimated on the order of $\pm 0.03 \pi$.}
\end{figure}

Remarkably, when scanning $\Delta \varphi$ in the upward and downward directions for a fixed excitation power we observe a bistable behavior, as predicted in Ref.~\cite{Petrov2013}.
In Fig.~\ref{fig3}(f), $\Delta \varphi$ is now decreased, starting from the situation $\Delta \varphi = 2 \pi$ from Fig.~\ref{fig3}(e). The expulsion or generation of single solitons takes place at different values of $\Delta \varphi$ than in the upward scan. In other words, there exist values of the phase difference between the beams, for which two different profiles --with either four or three solitons-- are stable: we evidence a bistability entirely controlled by the relative phase of the pumping beams.

The numerical simulation presented in Fig.~\ref{fig3}(g,h) is in good qualitative agreement with the measured phase scan, including the bistable behavior. There are however some differences: the theoretical patterns shown in the two panels transform into each other under the $\Delta\varphi \rightarrow 2\pi-\Delta\varphi$ transformation, while in the experiment, this symmetry is only approximately satisfied. Indeed the simulation shows a more regular displacement of the soliton pattern than the measurement.
For instance, when three solitons are stable, the measured pattern appears almost fixed in space for a wide range of $\Delta \varphi$. This can be explained by the presence of disorder in the wire, as confirmed by simulations when introducing a small potential dip to model a defect~\cite{Supplementary}. The slightly smaller bistability range observed in the experiments as compared to simulations could also be caused by disorder~\cite{Supplementary}, as well as by phase noise in the pump beams.

Figure~\ref{fig4} summarizes the measured number of solitons versus $\Delta \varphi$ in the upward and downward scans for different configurations of excitation powers and distances $d$. Abrupt switching between trains with $N$ and $N+1$ solitons is observed for $N$ ranging from 0 to 3. In each of these situations, we observe a well defined phase-controlled bistability.

In conclusion, we have demonstrated the ability to generate and control soliton trains in a 1D polariton quantum fluid. The ability to impose a controllable phase twist across the fluid using a coherent drive allows to reveal a novel bistable behavior. This experimental configuration offers a new perspective to explore the excitation spectrum of soliton trains in pump probe experiments.
Moreover, exploiting the polarization degree of freedom of polaritons, formation of spin domains~\cite{Petrov2013} and half soliton trains~\cite{Pinsker2014, Tercas2013, Tercas2014} have been predicted. Finally, from a more general perspective, as the response of a quantum fluid to a phase perturbation is quantitatively related to its superfluid fraction \cite{Pitaevskii2003, Carusotto2011, Janot2013}, our experiment opens the way to the experimental measurement of this quantity, crucial in the theory of driven dissipative quantum fluids~\cite{Wachtel2016}.

This work was supported by French National Research Agency (ANR) program Labex NanoSaclay projects Qeage (ANR- 11-IDEX-0003-02) and ICQOQS (ANR-10-LABX-0035), the French RENATECH network, the ERC grant Honeypol and the EU-FET Proactiv grant AQUS (Project No. 640800). IC is grateful to F. Dalfovo for insightful discussions.


%

\newpage

\clearpage

\section{Supplemental Material}

\subsection{I - Phase of the polariton fluid}

\begin{figure}[b]
	\includegraphics[width=\linewidth]{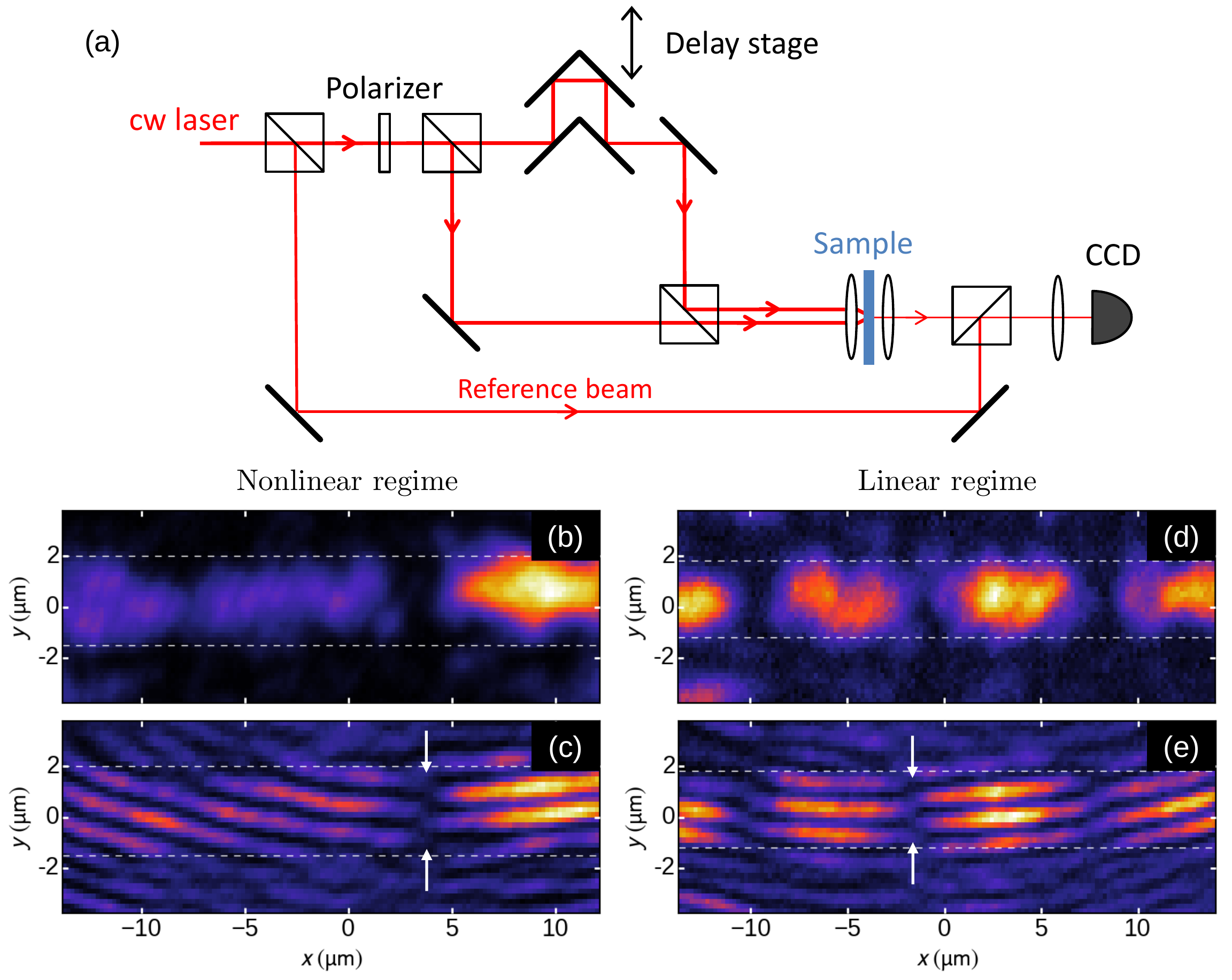}
	\caption{\label{figS1} (a) Sketch of the experimental setup used to measure the phase of the polariton fluid. (b) Real space emission profile and (c) corresponding interferogram of the wire, in the nonlinear regime. A phase jump of $\pi$ is clearly visible at the position of the soliton, indicated by the white arrows. (d) Real space emission profile and (e) corresponding interferogram in the linear regime. A phase jump of $\pi$ is also observed around each node of the standing wave pattern, see, e.g., the position indicated by the white arrows.
	}
\end{figure}

In contrast to ultracold atomic gases, it is possible to directly access the phase of the polariton fluid {\em in situ} by interferometric techniques. We use the pump laser beam, which has a constant phase, as a reference and we overlap this constant-phase reference beam with the real space emission from the wire (see Fig.~\ref{figS1}(a)). The phase of the polariton fluid can then be extracted from the resulting interferogram. In particular, a phase jump of $\pi$ can be detected in the interference pattern as a discontinuity in the fringes.
Fig.~\ref{figS1}(c) presents an interference pattern measured with a high pumping power for the polariton fluid, i.e., in the nonlinear regime. With the choice of experimental parameters here, a single soliton is present in the wire, as shown in Fig.~\ref{figS1}(b). A $\pi$ phase jump across the soliton is clearly evidenced in the interferogram, at the position indicated by the white arrows.

We emphasize, however, that measuring this phase jump of $\pi$ is not sufficient to ascertain that the density dip in the nonlinear regime is indeed a dark soliton. As mentioned in the main text, in the linear regime, a standing wave is formed between the pumping spots. Hence, also in the linear regime, there is a $\pi$ phase jump across each density dip, i.e., between each two nodes of the standing wave, as shown in Fig.~\ref{figS1}(d),(e), for example at the position indicated by the white arrows.
The nonlinear nature of the dark solitons present at high pumping power is confirmed mainly by their abrupt generation and expulsion in both power and phase scans, as shown in the main text. The presence of the $\pi$ phase jump is a mere sanity check.

\subsection{II - Soliton profile}

\begin{figure}[b]
	\includegraphics[width=0.8\linewidth]{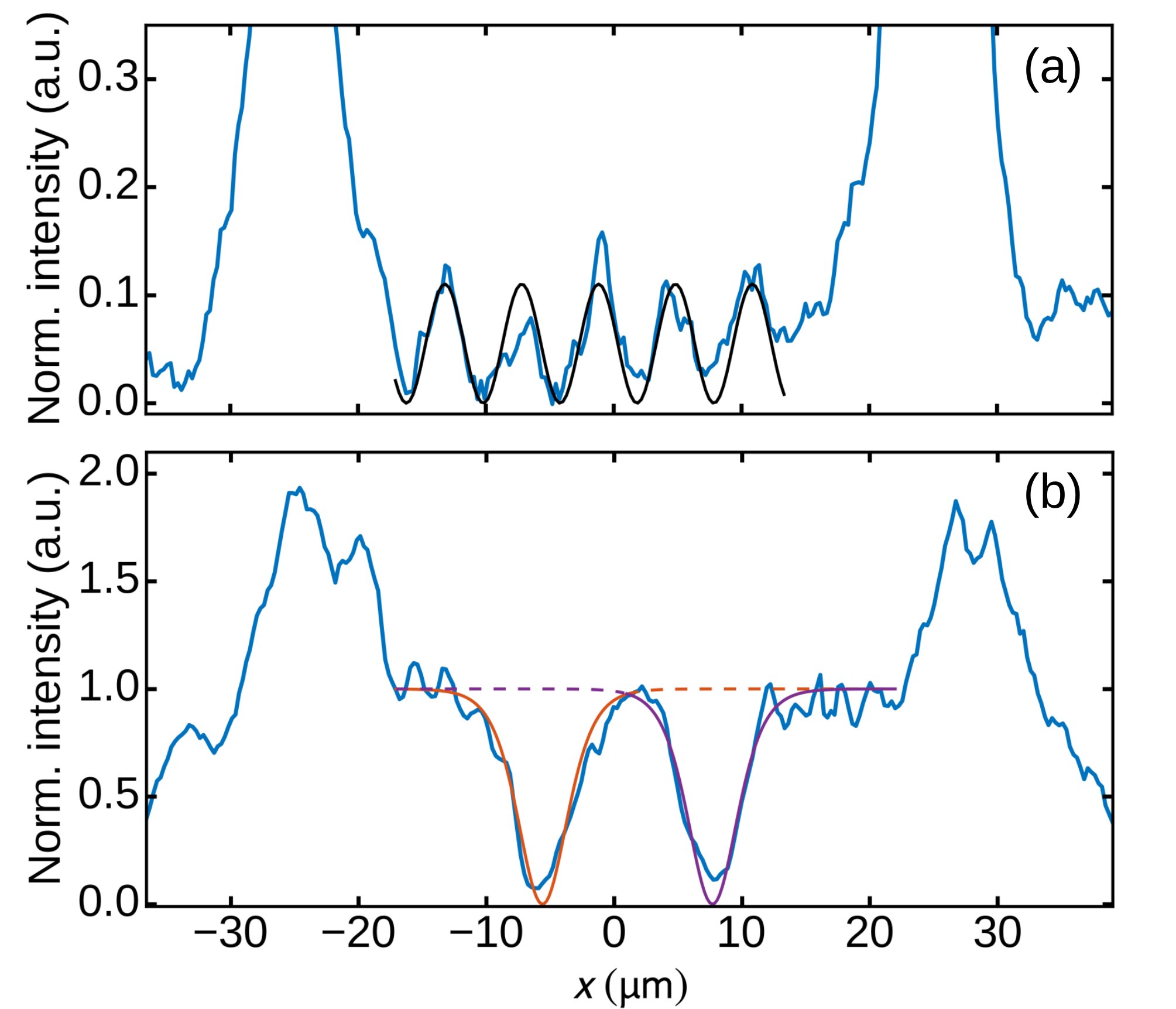}
	\caption{\label{figSnew} (a) Density profile in the linear regime, corresponding to Fig.~1(d) of the main text, fitted by a cosine function (black line). (b) Density profile in the nonlinear regime, corresponding to Fig.~2(b) of the main text. The orange (resp. purple) line is a fit of the left (right) soliton. Dashed lines indicate regions where the fits are not valid due to the presence of the second soliton.
	}
\end{figure}

\begin{figure*}[t]
	\includegraphics[width=\linewidth]{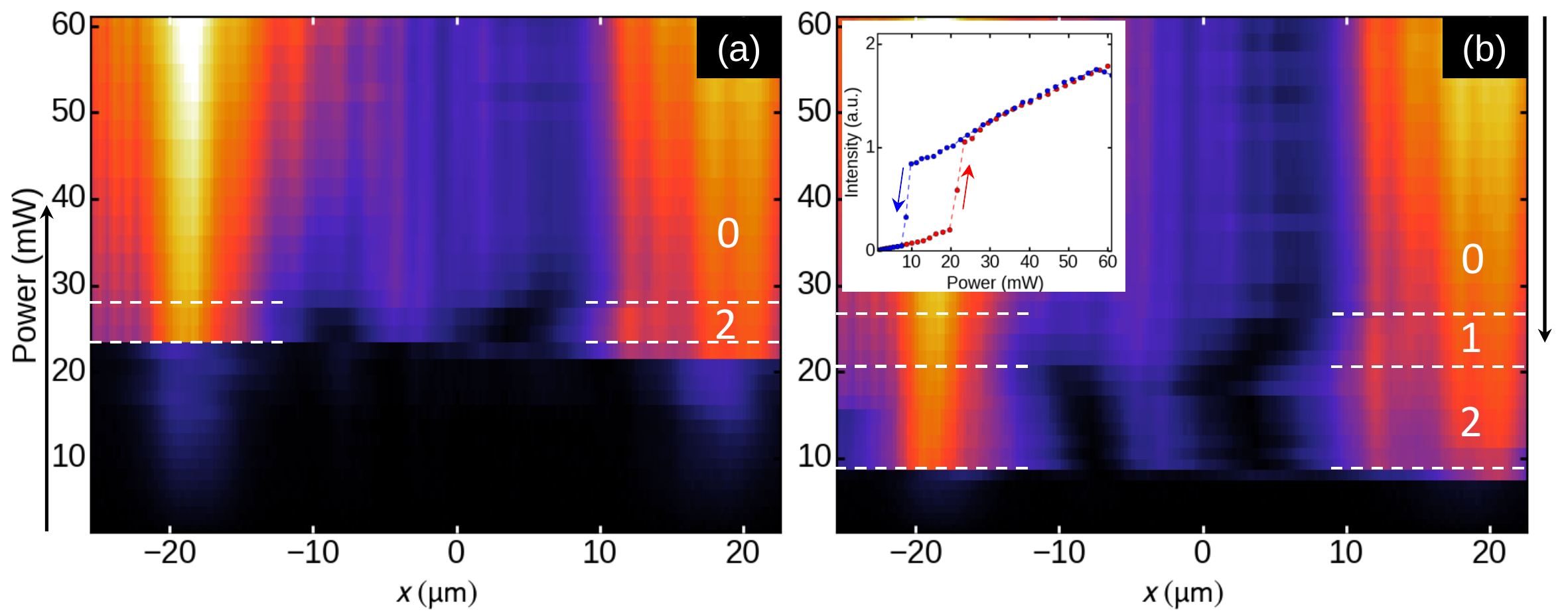}
	\caption{\label{figS2} (a) Integrated density profile measured as a function of pump power, for an increasing pump power. The number of solitons in the wire for a power range over which it remains constant is indicated in white. $\Delta E = 0.27 \ \mathrm{meV}$, $d = 40 \ \mathrm{\mu m}$ and $\Delta \varphi \approx 0$ (but non zero). (b) Power scan with the same parameters, but for a decreasing pump power. Inset: Total intensity in the wire for the upward (red dots) and downward (blue dots) pump power scans.
	}
\end{figure*}

As discussed in the main text, the solution of the Gross-Pitaevskii equation determining the profile of a soliton train is an elliptic function~\cite{Pitaevskii2003}. In the case of a profile with a single dark soliton, the density evolves as $n(x) = \mathrm{tanh}^2 \left((x - x_0)/\sigma\right)$, with $x_0$ the soliton position and $\sigma$ its half-width. In Fig.~\ref{figSnew}(b), we use this single soliton solution to fit the density profile in the nonlinear regime from Fig.~2(b) of the main text. The two soliton dips are fitted independently, using the same soliton width of $5.2 \ \mathrm{\mu m}$. The experimental data is well reproduced by the theoretical profile of two independent solitons in the region between the two excitation beams.

The fits reproduce the presence of the soliton dips on top of a flat background. The dips are separated by $13.5~\ \mathrm{\mu m}$, significantly more than twice their width. This situation is very different to what is observed in the linear regime, Fig.~\ref{figSnew}(a), in which a standard cosine-like interference pattern is observed with minima separated by exactly twice the FWHM.

These observations evidence that the density dips in the nonlinear regime are indeed solitons. Note that the nonlinear nature of the fluid is also confirmed by the behavior of the measured phase scans, characterized by sharp changes of the profile incompatible with the linear regime (compare Fig.~1(e) and Fig.~3(e) of the main text).

\subsection{III - Power-controlled bistability of the soliton train}

A resonantly pumped dissipative nonlinear system, such as a quantum well embedded in a microcavity, in the strong coupling regime, can exhibit bistability when scanning the pumping power upward or downward \cite{Baas2004b}. In this section, we present experiments revealing that such a bistable behavior is also present in our system. The bistability affects not only the transmitted intensity, but also the field profile in the cavity. We also report a bistability of the soliton pattern, controlled by the pumping power.

Fig.~\ref{figS2}(a) presents the evolution of the real space emission profile along the wire, integrated over the transverse direction, in an upward scan of the pumping power. Fig.~\ref{figS2}(b) is the corresponding downward scan, with the same experimental parameters and starting from the maximum pumping power reached in the upward scan. $\Delta \varphi$ is fixed to a value close to zero (but non zero). The threshold power corresponding to the transition from the linear to the nonlinear regime, measured at $P = 23 \ \mathrm{mW}$ in the upward scan, is lowered in the downward scan to $P = 9 \ \mathrm{mW}$. As one can see in the inset of Fig.~\ref{figS2}(b), this behavior corresponds to a standard hysteresis loop.

Focusing now on the evolution of the profile in the nonlinear regime, we notice, in the upward scan, a second transition $P = 28 \ \mathrm{mW}$ from a profile containing two solitons directly above threshold, to a profile with zero solitons for higher pumping powers. The mechanism responsible for such a transition has already been discussed in the main text (Fig.~2(e)). In the downward scan however, a profile that contains a single soliton is clearly visible from $P = 27 \ \mathrm{mW}$ down to $P = 21 \ \mathrm{mW}$, before the transition to a pattern with two solitons when the power is further decreased. Thus, a bistable behavior of the soliton pattern in the nonlinear regime is identified. Note that the existence of a regime with a single soliton is due to the fact that $\Delta \varphi$ is not exactly 0, which relaxes the parity condition for the polariton fluid. Profiles with an even number of solitons are nevertheless still more favorable, which is why the profile with a single soliton is not observed in the upward scan.

This power-controlled bistability is the counterpart of the phase-controlled bistability presented in the main text, in the sense that the number of solitons in the wire can be controlled by two independent parameters: the pumping power $P$, and the phase twist imposed across the wire, $\Delta \varphi$. Tuning either of these parameters, the expulsion or generation of a soliton is an abrupt event inherent to the discrete nature of the solitons, and additionally such a transition is associated with a hysteresis when scanning a single parameter.

\subsection{IV - Influence of disorder}

\begin{figure*}[t]
	\includegraphics[width=\linewidth]{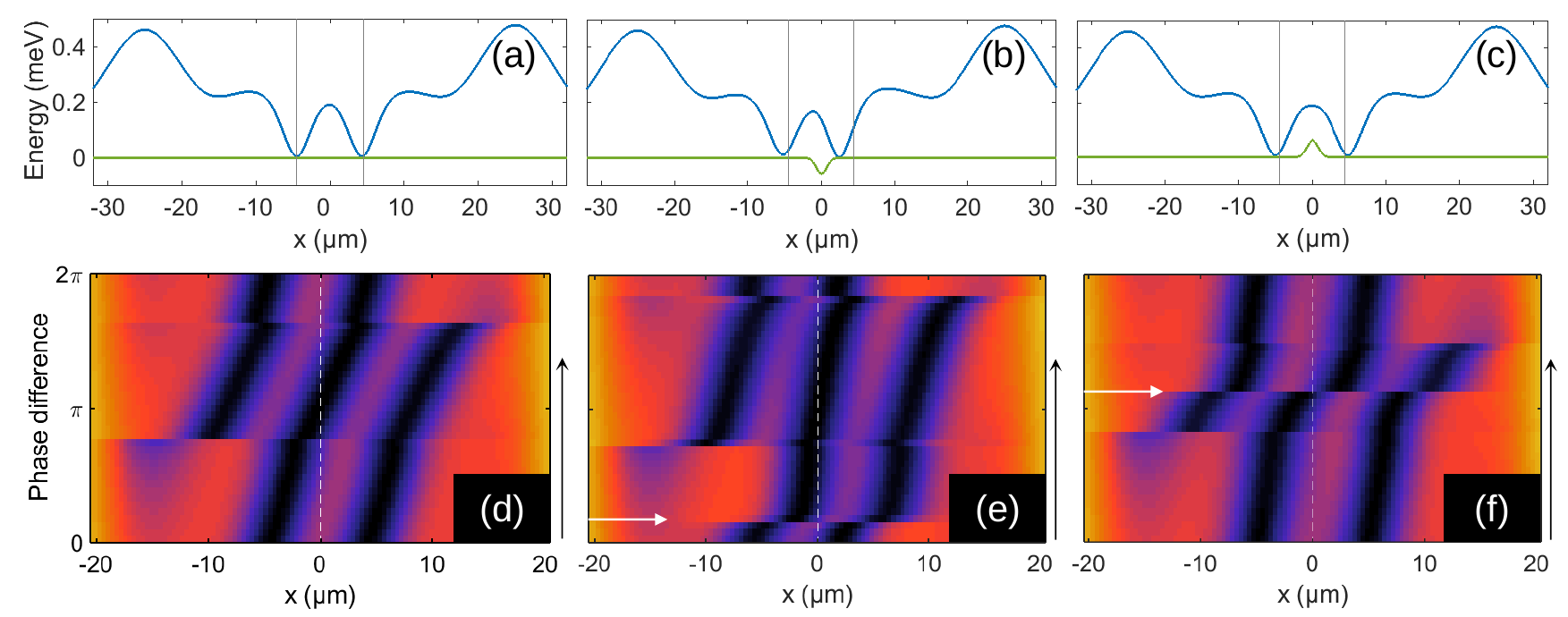}
	\caption{\label{figS3} Top row -- Numerical simulations of the density profile in a wire with $\Delta \varphi = 0$ and: (a) no defect. (b) $V_{def} = -60 \mathrm{\mu eV}$, $x_{def} = 0 \mathrm{\mu m}$. (c) $V_{def} = 60 \mathrm{\mu eV}$, $x_{def} = 0 \mathrm{\mu m}$. On each panel, the blue line is the local interaction energy, proportional to the local polariton density ($E_{int}(x) = \hbar g n(x)$). The green line is the potential energy arising from defects. Gray lines are guides for the eye, indicating the position of the solitons in the profile without defects. Bottom row -- Corresponding phase scans in the upward direction. The white dotted line is a guide for the eye, indicating the position of the defect in (e) and (f).
	}
\end{figure*}

In this section, we discuss the influence of disorder in the wire on the soliton train generated in the nonlinear regime.
We perform numerical simulations as described in the main text, including an additional potential energy term that accounts for the presence of defects in the wire. A defect is modeled by a gaussian potential:
\begin{equation}
V(x) = V_{def} e^{-\frac{(x-x_{def})^2}{w^2}},
\end{equation}
where the defect depth $V_{def}$ can be either positive or negative, $x_{def}$ is the defect position and $w$ its width.

From the general theory of solitons in atomic Bose gases~\cite{Konotop2004}, it is known that for repulsive $V_{def}>0$ defect potentials, the energy of a dark soliton is minimum when the soliton is located at the defect position where the background atomic density is lower (and viceversa for an attractive defect). This is easily understood in a perturbative picture as the interaction energy with the defect is proportional to the local particle density. On the other hand, the effective mass of a dark soliton seen as a quasiparticle is negative, as intuitively understood from the fact that the dark soliton corresponds to missing particles.
As a result, while energy minimization suggests that a dark soliton tends to bind to a repulsive defect, its actual kinematics is characterized by a repulsive acceleration~\cite{Frantzeskakis2002}.

Even if we are not aware of any complete theoretical study for polariton fluids, we can reasonably expect that these features remain valid also in this case. As the dissipative nature of these systems reduces the importance of energetic arguments, the physics is however likely to be dominated by the kinetic aspects and our simulations appear to confirm this naive expectation.

Fig.~\ref{figS3}(a-c) present the results of numerical simulations carried out with various defects (b,c), compared to a simulation without defects (a). The latter corresponds to the simulation of Fig.~2(c) of the main text. In both Fig.~\ref{figS3}(b) and (c), the defect width is $w = 1 \ \mathrm{\mu m}$. It is clearly visible that the presence of defects modifies the position of the solitons in the wire. More precisely, a negative defect has an attractive effect for a soliton, while a positive defect repels solitons.

Fig.~\ref{figS3}(d-f) shows numerical simulations of the evolution of the soliton pattern in a scan of $\Delta \varphi$, highlighting the effect of disorder on the soliton train position. 

As discussed in the main text, the displacement of the soliton pattern, shifted rightwards as $\Delta \varphi$ is increased, is homogeneous in the scan with no defects (d). 

In the case of a negative defect (e), the displacement is distorted: due to the attractive effect of the negative defect, a soliton coming close to the defect position is pinned and stays at the defect position in a finite range of $\Delta \varphi$. Beyond a threshold value of $\Delta \varphi$, the soliton is abruptly depinned as indicated by a white arrow in Fig.~\ref{figS3}(e). Such behavior is very close to the one experimentally observed and shown in Fig.3.(e),(f) of the main text.

For a positive defect on the other hand (f), there is no value of $\Delta \varphi$ for which a soliton is at the defect position, confirming the repulsive effect of the potential step. Moreover, because of this repulsion, an abrupt jump of the soliton pattern is observed, with a soliton jumping from the left of the defect to its right (indicated by the white arrow), when $\Delta \varphi$ is increased to the point that the solution with two solitons on the right of the defect becomes more stable than two solitons on the left.

This interpretation in terms of disorder is further supported by the observation that different sections of the sample showed slightly different soliton profiles while keeping the same excitation conditions. \\

Note that the values of $\Delta \varphi$ corresponding to the generation and expulsion of a soliton are affected by the presence of a defect. In particular, the hysteresis range measured during a phase scan depends significantly on the disorder in the wire. \\

\subsection{IV - Phase scans for various soliton trains}

\begin{figure}[b]
	\includegraphics[width=\linewidth]{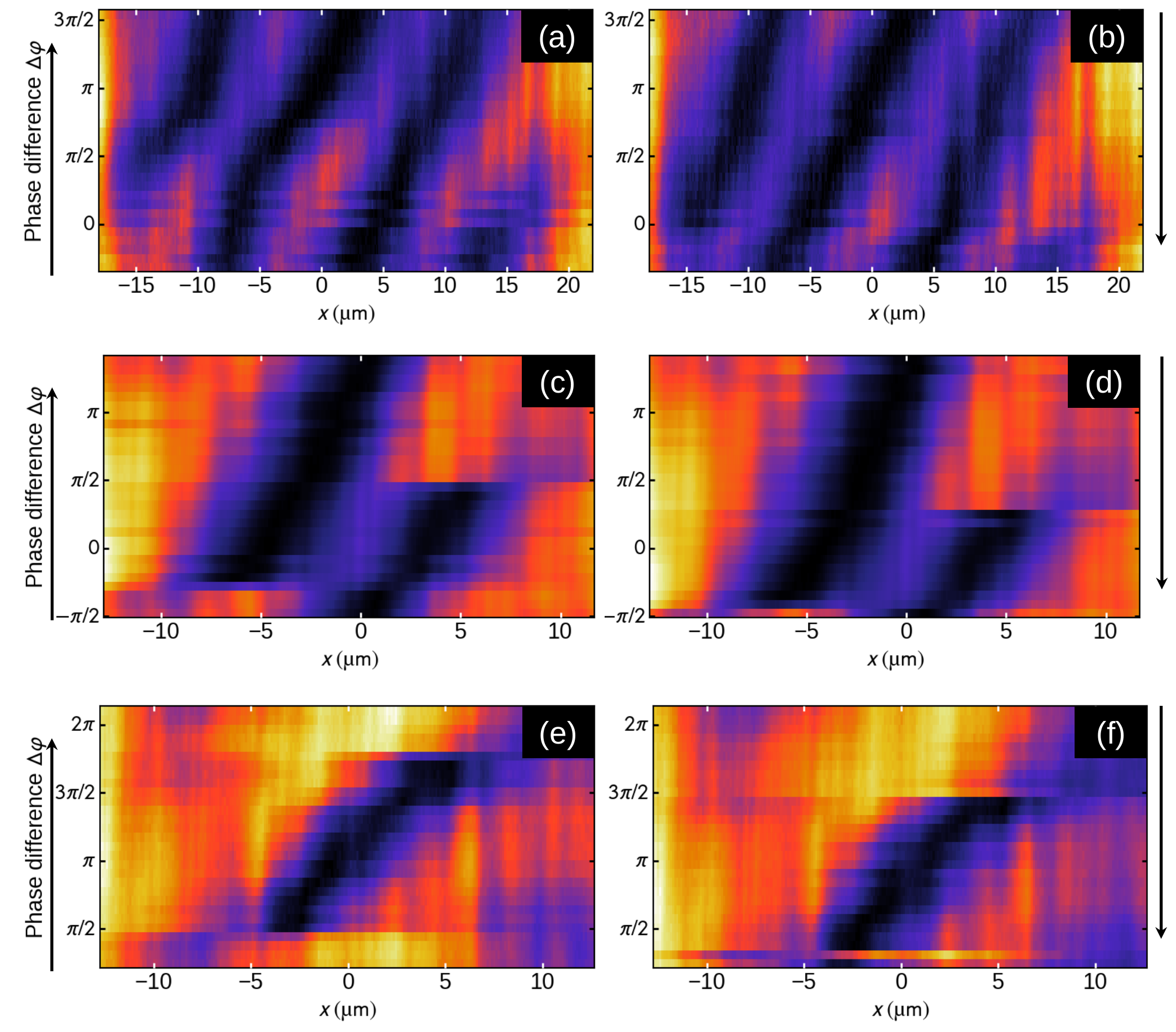}
	\caption{\label{figS4} Left column -- Phase scans in the upward direction ($\Delta \varphi$ increasing). Right column -- Corresponding phase scans in the downward direction. The parameters for each scans are: (a),(b) $\Delta E = 0.21 \ \mathrm{meV}$, $P = 42 \ \mathrm{mW}$, $d = 60 \ \mathrm{\mu m}$. (c),(d) $\Delta E = 0.35 \ \mathrm{meV}$, $P = 90 \ \mathrm{mW}$, $d = 40 \ \mathrm{\mu m}$. (e),(f) $\Delta E = 0.20 \ \mathrm{meV} $, $P = 103 \ \mathrm{mW} $, $d = 40 \ \mathrm{\mu m}$. Panels (a), (b) (resp. (c),(d) and (e),(f)) correspond to Fig. 4(b) (resp. (c) and (d)).
	}
\end{figure}

The number of solitons in the wire can be controlled by tuning the experimental parameters. Fig.~\ref{figS4} presents the phase scans, at constant pumping power, in the upward and downward direction, that correspond to the plots of Fig.~4 (b--d) of the main text. Parameters are indicated in the caption.

\end{document}